\documentclass[a4paper,11pt]{article}
\usepackage{pos}
\title{The hadronic contribution to the running of the electroweak gauge couplings}

\author*[a,b]{Alessandro Conigli}
\author[a,b]{Dalibor Djukanovic}
\author[c]{Georg von Hippel}
\author[d]{Simon Kuberski}
\author[a,c,d]{Harvey B. Meyer}
\author[e]{Kohtaroh Miura}
\author[c,f]{Konstantin Ottnad}
\author[g]{Andreas Risch}
\author[a,b,c]{Hartmut Wittig}
\affiliation[a]{Helmholtz Institute Mainz, Johannes Gutenberg-Universit\"{a}t Mainz, \\ 55099 Mainz, Germany}

\affiliation[b]{GSI Helmholtz Centre for Heavy Ion Research,\\ 64291 Darmstadt, Germany}

\affiliation[c]{PRISMA$^{++}$ Cluster of Excellence and Institut f\"{u}r Kernphysik, Johannes Gutenberg-Universit\"{a}t Mainz,\\ 55099 Mainz, Germany}

\affiliation[d]{Theoretical Physics Department, CERN, \\1211 Geneva 23, Switzerland
}

\affiliation[e]{Institute of Particle and Nuclear Studies, High Energy Accelerator Research Organization (KEK), \\ Tsukuba 305-0801, Japan}

\affiliation[f]{Helmholtz-Institut f\"{u}r Strahlen- und Kernphysik and Bethe Center for Theoretical Physics, Universit\"{a}t Bonn,  D-53115 Bonn, Germany}

\affiliation[g]{Department of Physics, University of Wuppertal, \\ Gaussstr. 20, 42119
	Wuppertal, Germany}

\emailAdd{aconigli@uni-mainz.de}

\abstract{
We present an updated determination of the hadronic vacuum polarization contribution to the running of the electromagnetic coupling $\Delta\alpha_{\mathrm{had}}^{(5)}(-Q^2)$, and of the electroweak mixing angle in the space-like momentum range up to $12 \ \mathrm{GeV}^2$. Using $N_f=2+1$ CLS ensembles at five values of the lattice spacing and several pion masses, including the physical point, we achieve a significantly enhanced precision over our previous result. A refined analysis strategy based on telescopic series and a new family of kernel functions enables a clean separation of distinct Euclidean regions, disentangling strong cutoff effects at short distances from the pronounced chiral dependence at larger ones. Employing the Euclidean split technique, we convert our lattice results into an ab initio estimate of $\Delta\alpha_{\mathrm{had}}^{(5)}(M_Z^2)$. A  comparison with results from other lattice calculations and phenomenology is performed. We also analyze improvement scenarios required to match the projected precision of future electroweak measurements at next-generation colliders.
\begin{flushright}
	MITP-26-016
	\\
	CERN-TH-2026-075
\end{flushright}
}

\FullConference{The 42nd International Symposium on Lattice Field Theory (LATTICE2025)\\
2-8 November 2025\\
Tata Institute of Fundamental Research, Mumbai, India\\}

\begin{document}
\maketitle

\section{Introduction}

Precision tests of the Standard Model (SM) rely on accurate determinations of the fundamental electroweak parameters. Among these, the electromagnetic coupling evaluated at the $Z$-boson pole, $\alpha(M_Z^2)$, plays a central role in global electroweak fits and in the interpretation of high-energy collider measurements. The dominant theoretical uncertainty in the determination of $\alpha(M_Z^2)$ arises from the hadronic contribution to the running of the electromagnetic coupling, $\Delta\alpha_{\rm had}^{(5)}$. This quantity encodes the effects of the Hadronic Vacuum Polarization (HVP) function and therefore represents a key non-perturbative input for precision electroweak physics. Improving its determination is essential in order to fully exploit the physics potential of future high-precision experiments.

Another key electroweak observable is the running of the weak mixing angle $\sin^2\theta_W(q^2)$, whose scale dependence is also governed by HVP effects. Precise measurements of $\sin^2\theta_W$ over a wide range of momentum scales provide complementary tests of the SM and sensitivity to possible contributions from physics beyond it. In particular, upcoming low-energy parity-violating experiments, such as P2~\cite{Becker:2018ggl} and MOLLER~\cite{MOLLER:2014iki}, aim to determine the weak mixing angle with unprecedented precision.

Traditionally, the hadronic contributions to the running of the electroweak couplings have been determined using dispersive approaches based on experimental measurements of the cross section for $e^+e^- \to {\rm hadrons}$~\cite{Keshavarzi:2018mgv, Davier:2019can,Jeger_yellow_rep,Jegerlehner:2019alphaQEDc19}. While these data-driven determinations have reached impressive precision, they are affected by tensions between different experimental datasets, motivating independent first-principles calculations.

Lattice QCD provides a framework to compute the HVP non-perturbatively from first principles. In recent years, significant progress, driven in part by efforts to understand the muon $g-2$ \cite{Aliberti:2025beg}, has greatly improved lattice determinations relevant for precision observables, including the running of the electroweak couplings.

Preliminary results from this project were reported in~\cite{Conigli:2025tbb}, while the final analysis was presented in~\cite{companion_paper}. The latter significantly improves upon our previous calculation~\cite{Ce:2022eix} and achieves roughly twice the precision of recent phenomenological estimates~\cite{Davier:2017zfy,Keshavarzi:2018mgv,Davier:2019can,Keshavarzi:2019abf,Jeger_yellow_rep,Jegerlehner:2019alphaQEDc19}. In these proceedings we summarize the lattice QCD determination of the hadronic contributions to the running of the electromagnetic coupling and the electroweak mixing angle, and discuss improvement scenarios relevant for future collider programs.

\section{Hadronic Contributions to Electroweak Running}

The electromagnetic coupling $\alpha = e^2/(4\pi)$ is defined in the
Thomson limit. In the on-shell renormalization scheme its value at a
momentum scale $q^2$ is expressed as
\begin{equation}
	\alpha(q^2)=\frac{\alpha}{1-\Delta\alpha(q^2)},
\end{equation}
where $\Delta\alpha(q^2)$ describes the energy dependence of the coupling.  While the
leptonic contribution can be computed perturbatively with very high
precision, the hadronic contribution is dominated by
non-perturbative QCD effects at low energies. It can be expressed in terms of the subtracted HVP function $\bar\Pi(q^2)$ as
\begin{equation}
	\Delta\alpha_{\rm had}(q^2)=4\pi\alpha\,\mathrm{Re}\,\bar{\Pi}(q^2), 
	\qquad
	\bar{\Pi}(q^2)=\Pi(q^2)-\Pi(0).
\end{equation}

HVP effects also govern the running of the electroweak mixing angle $\sin^2\theta_W(q^2)$. The corresponding hadronic contribution can be written as
\begin{equation}
	(\Delta\sin^2\theta_W)_{\rm had}(q^2)
	=
	\Delta\alpha_{\rm had}(q^2)-\Delta\alpha_{2,{\rm had}}(q^2)
	=
	-\frac{4\pi\alpha}{\sin^2\theta_W(0)}\bar{\Pi}^{(Z,\gamma)}(q^2),
\end{equation}
where $\bar{\Pi}^{(Z,\gamma)}$ denotes the mixed electromagnetic--weak HVP function.

On the lattice the HVP is obtained from Euclidean two-point correlation functions of vector currents. In this work we employ the Time-Momentum Representation (TMR)~\cite{Bernecker:2011gh}, in which the subtracted polarization function is computed from the spatially summed vector correlator.
By computing the correlation functions of the electromagnetic and neutral weak currents, one obtains the HVP functions $\bar{\Pi}^{(\gamma,\gamma)}$ and $\bar{\Pi}^{(Z,\gamma)}$ entering the hadronic contributions to the running of $\alpha$ and $\sin^2\theta_W$. A detailed description of the theoretical framework and lattice methodology can be found in Ref.~\cite{companion_paper}.
	
\section{Computational strategy}\label{sec:strategy}

Our computation is based on CLS gauge ensembles with $N_f=2+1$ flavours of non-perturbatively $O(a)$-improved Wilson fermions and a L\"{u}scher-Weisz  gauge action  \cite{Bruno:2014jqa,Bali:2016umi, Mohler:2017wnb, Mohler:2020txx,Kuberski:2023zky}. We analyze an extended set of ensembles with five lattice spacings in the range $a\simeq 0.039\text{--}0.085\,\mathrm{fm}$ and pion masses down to (and including) the physical point, allowing for controlled chiral and continuum extrapolations. Further details of the ensemble set and analysis setup are given in Ref.~\cite{companion_paper}.

A key element of our strategy is a telescopic (window-like) decomposition of the subtracted HVP that  separates Euclidean regimes with different dominant systematics. Following the philosophy of short/intermediate/long-distance window methods introduced in the context of the muon $g\!-\!2$~\cite{RBC:2018dos}, we compute the total subtracted HVP as\footnote{We define $Q^2=-q^2$, such that $Q^2>0$ denotes space-like virtualities.}
\begin{equation}
	\bar{\Pi}(-Q^2)=\widehat{\Pi}(-Q^2)+\widehat{\Pi}(-Q^2/4)+\bar{\Pi}(-Q^2/16),
	\qquad
	\widehat{\Pi}(-Q^2)\equiv \Pi(-Q^2)-\Pi(-Q^2/4).
	\label{eq:hvp_split_proc}
\end{equation}
We refer to the three contributions in Eq.~\eqref{eq:hvp_split_proc} as high-, mid- and low-virtuality (HV/MV/LV) terms. This separation isolates the short-distance dominated contribution, where discretization effects are most pronounced, from the long-distance dominated piece, where chiral dependence and statistical noise play a larger role. The decomposition therefore allows us to disentangle systematics and to tailor the extrapolation strategy to the characteristic behaviour of each term; it also enables us to reach higher Euclidean momenta reliably, which is beneficial for matching to perturbative QCD in the determination at the $Z$ pole.

To further control short-distance cutoff effects, we additionally employ a kernel-subtraction strategy introduced in Ref.~\cite{Kuberski:2024bcj} and applied also in \cite{Beltran:2026ofp}. The idea is to modify the TMR kernels by subtracting a term whose contribution can be evaluated in continuum perturbation theory, thereby suppressing higher-order lattice artefacts and cancelling potentially log-enhanced cutoff effects in the very short Euclidean-time region~\cite{Ce:2021xgd,Sommer:2022wac}.  The full implementation and the specific strategy adopted for each individual channel  are described in Ref.~\cite{companion_paper}.
	
\section{Lattice results}\label{sec:lat_results}
	Our lattice results are obtained after extrapolating all contributions to the physical point in the isospin-symmetric theory and correcting for the dominant finite-volume effects. The full analysis strategy, including all channel-specific ingredients and extensive stability tests, is documented in Ref.~\cite{companion_paper}. For the physical-point extrapolation we use the dimensionless quark mass proxies
	\begin{equation}
		\phi_2 = 8t_0 m_\pi^2, \qquad \phi_4 = 8t_0\!\left(m_K^2+\tfrac12 m_\pi^2\right),
	\end{equation}
	and parametrize cutoff effects in terms of $X_a^2=a^2/(8t_0)$, with the scale set through the gradient-flow observable $t_0$~\cite{Luscher:2010iy} using $t_0^{\rm phys}$ from Ref.~\cite{Bussone:2025wlf}. The chiral and continuum limits are taken simultaneously using fit ans\"atze guided by Symanzik effective theory, including leading $O(a^2)$ effects and higher-order terms. 
	
		\begin{figure}
		\centering
		\includegraphics[scale=0.29]{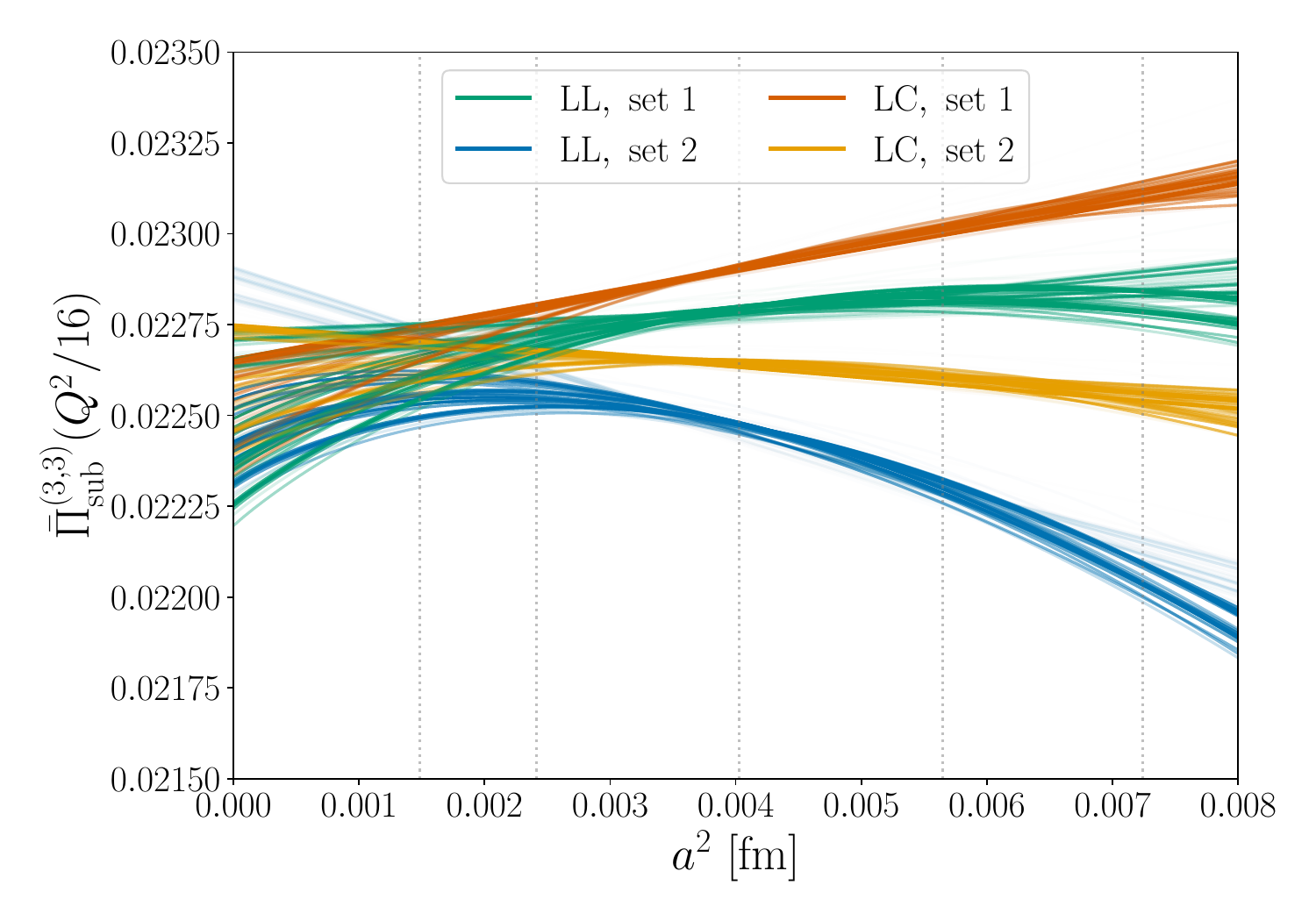}
		\includegraphics[scale=0.29]{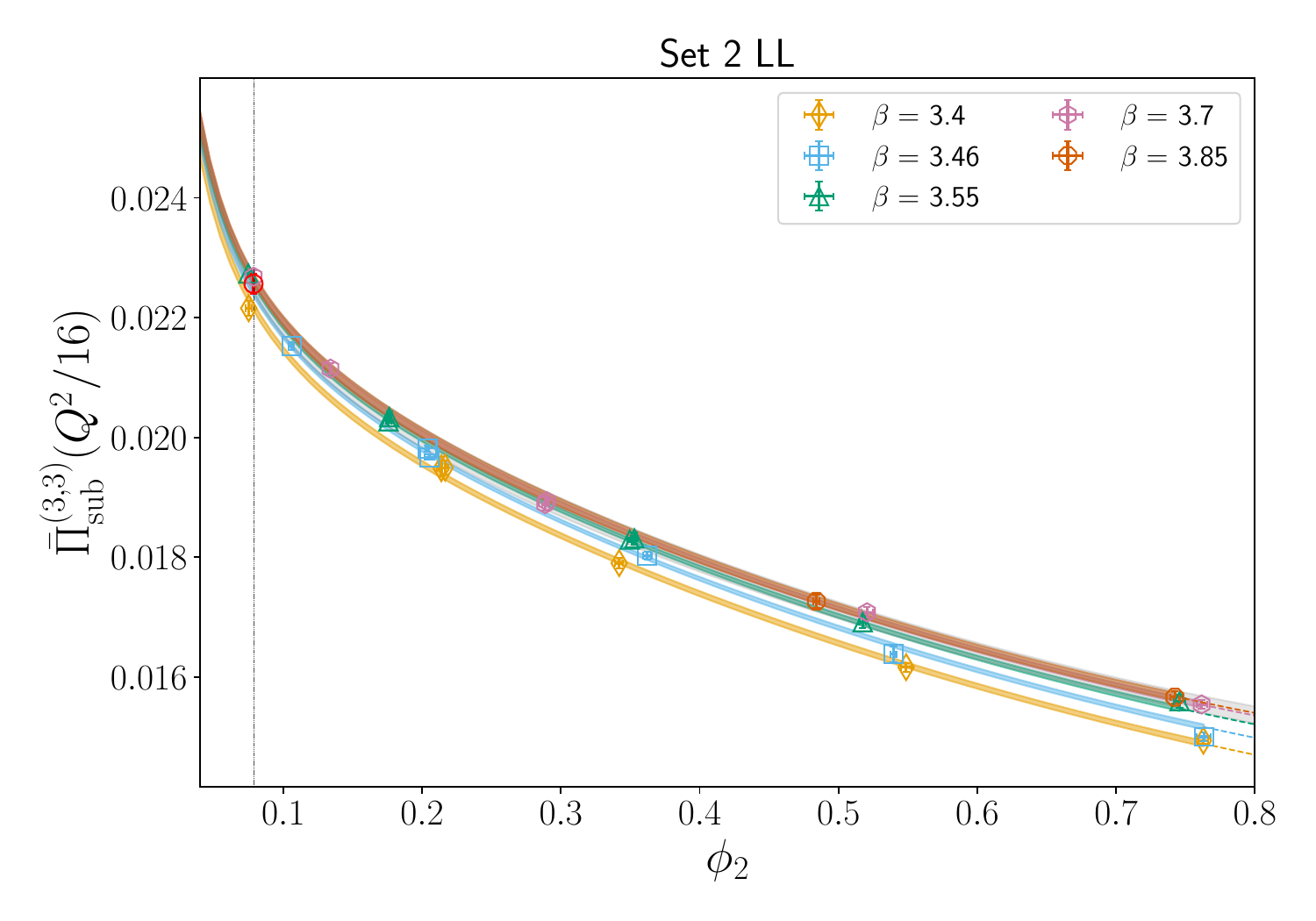}
		\caption{Illustration of the fits to the isovector contribution $\bar\Pi^{(3,3)}_{\mathrm{sub}}(Q^2/16)$ in the LV region. Left: continuum extrapolation for different improvement schemes and vector-current discretizations. Lines correspond to individual fits weighted by the model averaging procedure. Right: chiral extrapolation to the physical pion mass for the highest-weight fit, showing finite-$a$ chiral trajectories and the continuum dependence on $\phi_2$. Results are shown for $Q^2=9~\mathrm{GeV}^2$.}
		\label{fig:hv_isov_extrap_proc}
	\end{figure}

\begin{figure}
		\centering
         \includegraphics[scale=0.23]{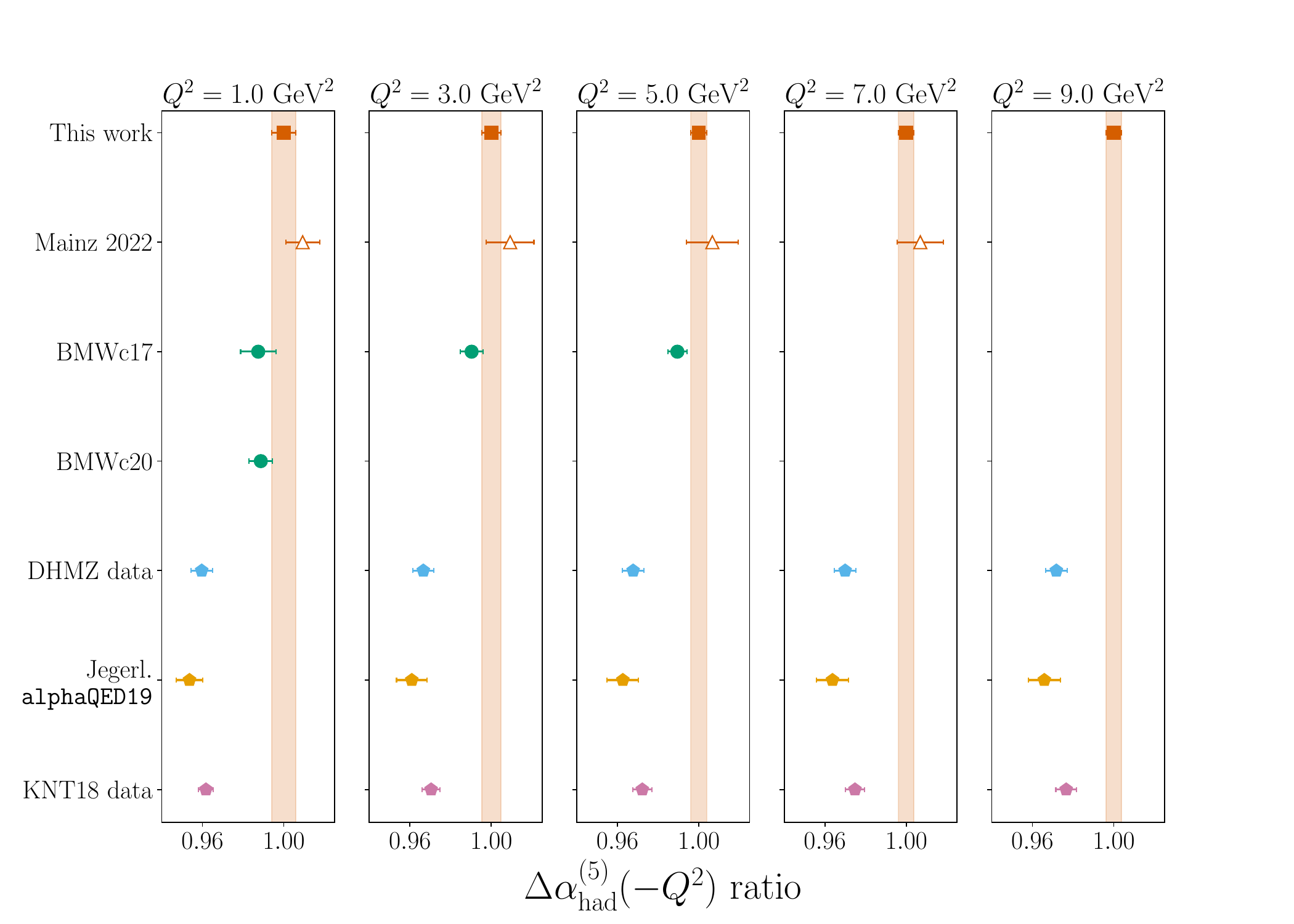}
       	\includegraphics[scale=0.235]{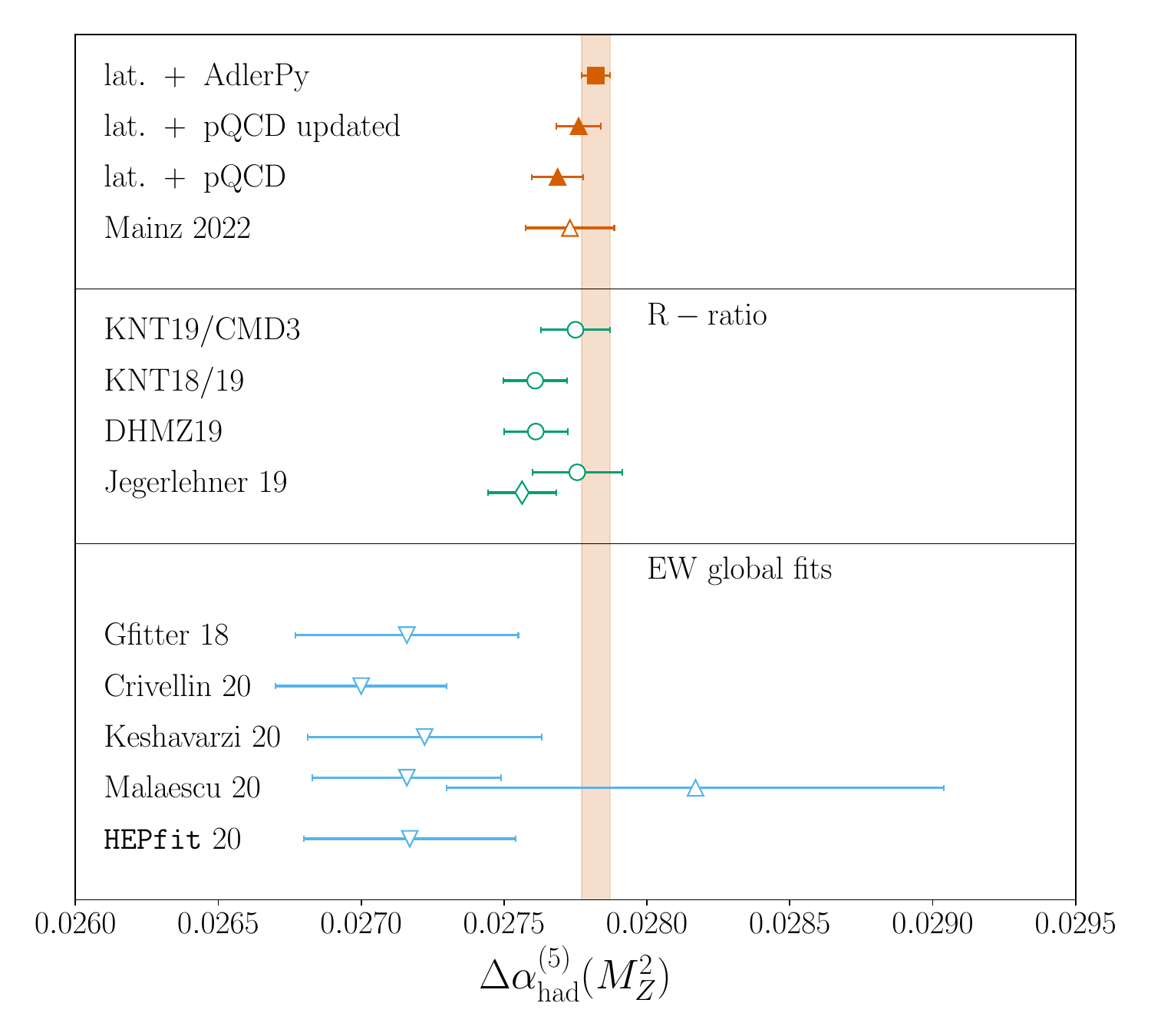}
		\caption{Summary of results for the hadronic running of the electromagnetic coupling. 
			Left: ratios of lattice and phenomenological determinations of 
			$\Delta\alpha_{\mathrm{had}}^{(5)}(-Q^2)$ to our central values; the orange band shows our total uncertainty, including the bottom-quark contribution. 
			Right: comparison of determinations of $\Delta\alpha_{\mathrm{had}}^{(5)}(M_Z^2)$. 
			Our main result (AdlerPy) and the two \texttt{pQCDAdler} variants are shown together with the Mainz 2022 result~\cite{Ce:2022eix}, dispersive evaluations based on the $R$-ratio (green), and values from global EW fits (blue).}
		\label{fig:comparison_spacelike_proc}
	\end{figure}

	The chiral dependence is found to differ significantly among the HV, MV and LV contributions: it is relatively mild in the short-distance dominated HV region, while it becomes increasingly important in the LV region, where long-distance physics gives the dominant contribution. Systematic uncertainties from the extrapolation are assessed by varying the fit forms and by applying dataset cuts (e.g.\ excluding the coarsest lattice spacing and/or heavier-pion-mass ensembles). Our quoted central values and systematic errors are obtained from a weighted model average~\cite{Jay:2020jkz}, with weights based on the Akaike Information Criterion (AIC)~\cite{Bruno:2022mfy}. Statistical error propagation is performed using the $\Gamma$-method~\cite{Wolff:2003sm} as implemented in the  \texttt{ADerrors} package~\cite{Ramos:2018vgu}.  An illustration of the extrapolations for the LV isovector contribution is shown in Fig.~\ref{fig:hv_isov_extrap_proc}.
	\\ \indent
	A major limitation of HVP calculations is the rapid degradation of the signal-to-noise ratio at large Euclidean times. To mitigate this effect we employ several noise-reduction techniques. For the dominant light-quark connected contribution we use Low-Mode Averaging (LMA)~\cite{Giusti:2004yp,DeGrand:2004qw} following Ref.~\cite{ Borsanyi:2020mff, Djukanovic:2024cmq}. We also apply the bounding method~\cite{RBC:2018dos,Gerardin:2019rua,Borsanyi:2016lpl,Lehner2016} to control the long-distance tail of the correlators, and in selected ensembles we reconstruct the correlator using dedicated spectroscopy studies~\cite{Gerardin:2019rua,Djukanovic:2024cmq}. These techniques substantially improve the precision at low virtualities.
	\\ \indent
	Finite-volume effects are corrected using the hybrid strategy developed in Refs.~\cite{Ce:2022eix,Djukanovic:2024cmq}. In the isovector channel we combine the Hansen--Patella approach~\cite{Hansen:2019rbh,Hansen:2020whp} at short Euclidean times with the Meyer--Lellouch--L\"uscher formalism~\cite{Meyer:2011um} for the long-distance regime. Following \cite{Borsanyi:2020mff}, ensembles are first corrected to a common reference value of $m_\pi L$ before the chiral--continuum extrapolation, and the remaining correction to infinite volume is applied in the continuum limit. Further details are given in Ref.~\cite{companion_paper}.
	\\ \indent
	Isospin-breaking (IB) effects are subleading for the observables considered here. We estimate them using lattice QCD+QED calculations together with a phenomenological model adapted from studies of $a_\mu^{\rm hvp}$~\cite{Biloshytskyi:2022ets,Parrino:2025afq,Erb:2025nxk}, evolved perturbatively to higher virtualities~\cite{Kataev:1992dg}. Their impact is small compared to the total uncertainty.
	\\ \indent
	Our main results in the spacelike region, for
	$0.25~{\rm GeV}^2 \le Q^2 \le 12~{\rm GeV}^2$, are the determinations of
	$\Delta\alpha_{\rm had}^{(5)}(-Q^2)$ and
	$(\Delta\sin^2\theta_W)_{\rm had}(-Q^2)$.
	A comparison of our $\Delta\alpha_{\rm had}^{(5)}(-Q^2)$ results with other lattice and dispersive evaluations is shown in the left panel of Fig.~\ref{fig:comparison_spacelike_proc}. Our updated results are consistent with Ref.~\cite{Ce:2022eix} within uncertainties, while mild discrepancies  are observed with BMW results~\cite{Budapest-Marseille-Wuppertal:2017okr,Borsanyi:2020mff}. Dispersive evaluations~\cite{Keshavarzi:2018mgv,Davier:2019can,Jeger_yellow_rep,Jegerlehner:2019alphaQEDc19} lie systematically below lattice determinations. A detailed discussion is given in Ref.~\cite{companion_paper}.

	\section{Running of $\alpha$ to the $Z$ pole}\label{sec:z_pole}
	
We convert our lattice determination of the space-like running,
$\Delta\alpha_{\mathrm{had}}^{(5)}(-Q^2)$, into an estimate of the
hadronic contribution at the $Z$ pole,
$\Delta\alpha_{\mathrm{had}}^{(5)}(M_Z^2)$, an important input for
precision electroweak analyses
\cite{Jeger_yellow_rep,Jegerlehner:2011mw,Baak:2014ora}. To connect the
Euclidean region accessible in lattice QCD with the time-like scale
$M_Z^2$, we employ the Euclidean split technique
\cite{Eidelman:1998vc,Jegerlehner:2008rs}. This method separates the
total running into a non-perturbative lattice contribution at a
matching scale $Q_0^2$, a perturbative evolution from $Q_0^2$ to
$M_Z^2$, and a small perturbative continuation from $-M_Z^2$ to
$+M_Z^2$:
\begin{equation}\label{eq:euclidean_splitting_proc}
	\begin{split}
		\Delta\alpha_{\mathrm{had}}^{(5)}(M_Z^2)
		&=
		\Delta\alpha_{\mathrm{had}}^{(5)}(-Q_0^2)
		\\ &\quad
		+\Big[
		\Delta\alpha_{\mathrm{had}}^{(5)}(-M_Z^2)
		-
		\Delta\alpha_{\mathrm{had}}^{(5)}(-Q_0^2)
		\Big]
		+\Big[
		\Delta\alpha_{\mathrm{had}}^{(5)}(M_Z^2)
		-
		\Delta\alpha_{\mathrm{had}}^{(5)}(-M_Z^2)
		\Big].
	\end{split}
\end{equation}
	The first term is taken from our lattice determination of the
	space-like HVP. The last term is numerically
	small and is estimated in pQCD as \cite{Jeger_yellow_rep}
	$
		\Delta\alpha_{\mathrm{had}}^{(5)}(M_Z^2)
		-
		\Delta\alpha_{\mathrm{had}}^{(5)}(-M_Z^2)
		=
		0.000\,045(2).
	$	
	The remaining ingredient is the perturbative running between $Q_0^2$
	and $M_Z^2$, obtained from the Adler function $D(Q^2)$. We evaluate
	this contribution using three perturbative implementations: the
	original \texttt{pQCDAdler} setup used in Mainz 2022
	\cite{Ce:2022eix}, an updated version with FLAG24 inputs
	\cite{FlavourLatticeAveragingGroupFLAG:2024oxs}, and the public
	package \texttt{AdlerPy}~\cite{Hernandez:2023ipz}, which we adopt as
	our preferred estimate. The three approaches give compatible results
	for $Q_0^2 \gtrsim 3~\mathrm{GeV}^2$, while lower matching scales show
	the expected loss of perturbative stability.
	
	In the right panel of Fig.~\ref{fig:comparison_spacelike_proc} we
	compare our result for
	$\Delta\alpha_{\mathrm{had}}^{(5)}(M_Z^2)$ with these variants, with
	our previous Mainz 2022 determination \cite{Ce:2022eix}, and with
	 phenomenological evaluations. While sizeable tensions with data-driven estimates
	are observed in the space-like region, they are significantly reduced at the
	$Z$~pole once the perturbative running is included. Our preferred
	result is slightly larger than standard dispersive evaluations,
	typically at the $1$--$2\,\sigma$ level, while remaining compatible
	with determinations based on CMD-3 data
	\cite{DiLuzio:2024sps,CMD-3:2023alj,CMD-3:2023rfe}.

\section{Future improvement scenarios}\label{sec:future}

Future $e^+e^-$ colliders will require significantly improved
precision in the determination of the running of the electromagnetic
coupling. In particular, the FCC-ee aims at an absolute precision of
$(3$--$5)\times10^{-5}$ on $\alpha(M_Z^2)$ from measurements of the
muon forward--backward asymmetry
\cite{Janot:2015gjr,Freitas:2019bre,Jeger_yellow_rep}. This translates
into a required $(1.0$--$1.7)$\,\textperthousand\ precision on the
hadronic contribution, with the upper bound already comparable to the
uncertainty achieved in this work. Moreover, a recent proposal based
on $Z$-pole measurements could reach a projected statistical
sensitivity below $10^{-5}$~\cite{Riembau:2025ppc}. These targets
motivate further improvements in the theoretical determination of
$\Delta\alpha_{\mathrm{had}}^{(5)}(M_Z^2)$.

In our approach, the running is obtained by combining the lattice
determination of $\Delta\alpha_{\mathrm{had}}^{(5)}(-Q_0^2)$ with a
perturbative evolution from the matching scale $Q_0^2$ to the
$Z$ pole. The final uncertainty therefore receives contributions from
both the lattice calculation and the perturbative running.

These two sources of uncertainty exhibit opposite behaviour as the
matching scale is varied. Increasing $Q_0^2$ reduces the interval over
which perturbative QCD is applied, thereby decreasing the perturbative
uncertainty. At the same time, larger Euclidean momenta are more
challenging on the lattice, where discretization effects become more
important. The optimal matching scale must therefore balance these
competing effects.

To explore possible improvement scenarios, we construct a simple model
for the projected total uncertainty,
\begin{equation}
	\sigma_{\mathrm{tot}}^2(Q_0^2)=
	\sigma_{\mathrm{pQCD}}^2(Q_0^2)+
	\sigma_{\mathrm{lat}}^2(Q_0^2),
\end{equation}
where the perturbative term depends on the precision of the theory
inputs entering the Adler function, in particular $\alpha_s(M_Z)$ and
the heavy-quark masses, while the lattice term reflects the
uncertainty of $\Delta\alpha_{\mathrm{had}}^{(5)}(-Q_0^2)$.

\begin{figure}[t]
	\centering
	\includegraphics[scale=0.38]{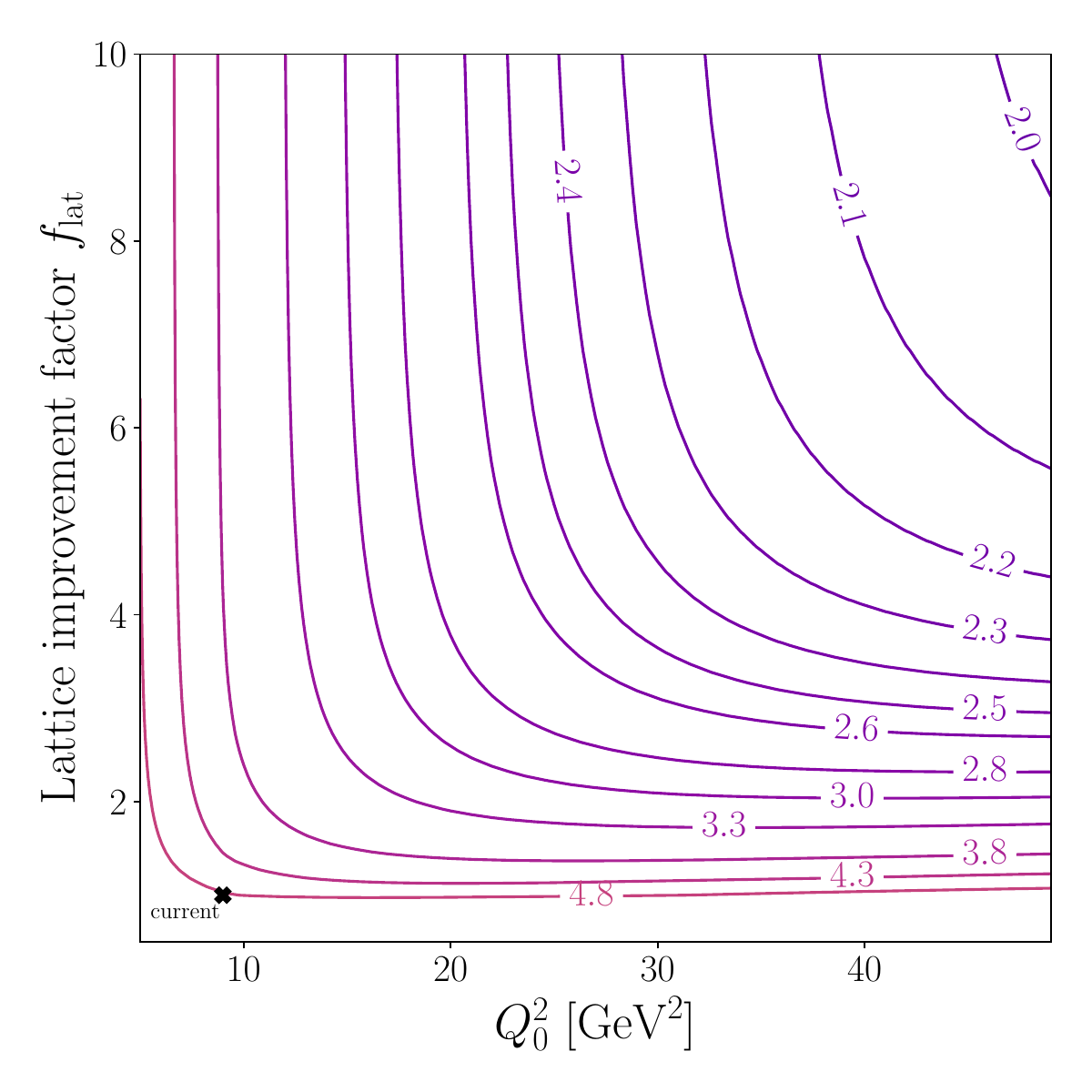}
	\caption{Projected total uncertainty $\sigma_{\mathrm{tot}}$ in the $(Q_0^2,f_{\mathrm{lat}})$ plane. Contours show the total error (in units of $10^{-5}$) as a function of the matching scale and of a global lattice improvement factor that rescales the current lattice uncertainty. The black cross indicates the uncertainty achieved in this work.}
	\label{fig:mom_flat_plot}
\end{figure}

The resulting projections are illustrated in
Fig.~\ref{fig:mom_flat_plot}, which shows contour lines of the total
uncertainty as a function of the matching scale and of a global
lattice improvement factor. The current determination corresponds to
the point $Q_0^2=9~\mathrm{GeV}^2$. The figure indicates that neither
increasing the matching scale alone nor improving the lattice
precision at fixed $Q_0^2$ leads to substantial gains. Instead, the
most efficient reduction of the uncertainty is achieved by combining
moderate lattice improvements with a moderately larger matching
scale.

In particular, reaching a target precision of
$\sim 3\times10^{-5}$, relevant for future electroweak measurements,
would require roughly a factor two reduction of the lattice
uncertainty together with a matching scale around
$Q_0^2\sim20~\mathrm{GeV}^2$. Improvements in the perturbative inputs
play a secondary role, with the uncertainty in $\alpha_s$ providing
the dominant contribution among them. Overall, continued advances in
lattice QCD, together with improved determinations of key Standard
Model parameters, will be essential to further reduce the uncertainty
in $\Delta\alpha_{\mathrm{had}}^{(5)}(M_Z^2)$.
	
\section{Conclusions}

We have presented a high-precision lattice QCD determination of the
hadronic contribution to the running of the electromagnetic coupling
and of the electroweak mixing angle over a broad range of space-like
momenta up to $Q^2=12~\mathrm{GeV}^2$. The calculation benefits from
significantly improved statistical precision and control of systematic
effects, enabled by an expanded ensemble set and a refined
chiral--continuum extrapolation strategy based on the telescopic
window decomposition of the HVP correlator.

Our results for $\Delta\alpha_{\mathrm{had}}^{(5)}(-Q^2)$ confirm a
persistent tension with phenomenological determinations based on the
$R$-ratio in the space-like region. Combining the lattice results with
the perturbative running through the Euclidean split technique yields
a determination of
$\Delta\alpha_{\mathrm{had}}^{(5)}(M_Z^2)$ with a total relative
uncertainty of about $1.7$\,\textperthousand, improving the precision
of current phenomenological estimates by roughly a factor two and
reducing the dominant non-perturbative uncertainty entering
electroweak precision tests.

We have also investigated future improvement scenarios, showing that
moderate reductions of lattice uncertainties together with an
extension of the accessible momentum range would allow the precision
required by future electroweak measurements to be reached. Further
progress will involve extending the lattice calculation to higher
virtualities, incorporating full isospin-breaking effects, and
exploring alternative strategies to reduce discretization effects at
large momenta.

\paragraph{Acknowledgements}
\small
We thank Andrew Hanlon, Nolan Miller, Ben H\"{o}rz, Daniel Mohler, Colin Morningstar and Srijit Paul for the collaboration on the data generation and analysis for the spectral reconstruction.
We thank Volodymyr Biloshytskyi for providing auxiliary data used in the 
estimate of isospin-breaking effects.  
A.C. is grateful to Arnau Beltran for valuable discussions and crosschecks performed throughout the analysis.
We are grateful to Marco Cè for sharing data used to produce comparison plots.
We are grateful  to Bogdan Malaescu and the authors of refs. \cite{Davier:2019can} for sharing their data for the running of $\alpha$ at $Q^2>0$.
We thank Alex Keshavarzi and the authors of refs. \cite{Keshavarzi:2018mgv} for sharing the $R$-ratio data with covariance matrix, which were used to calculate tabulated data for the running of $\alpha$ at $Q^2>0$.
We are grateful to our colleagues in the CLS initiative for sharing ensembles.
Calculations for this project were performed on the HPC
clusters Clover and HIMster-II at the Helmholtz Institute Mainz and
Mogon-II and Mogon-NHR at Johannes Gutenberg-Universität (JGU)
Mainz, as well as on the GCS Supercomputers JUQUEEN and JUWELS at 
the Jülich Supercomputing Centre (JSC), HAZELHEN and HAWK at the 
Höchstleistungsrechenzentrum Stuttgart (HLRS), and SuperMUC at the 
Leibniz Supercomputing Centre (LRZ).
The authors gratefully acknowledge the support of the Gauss Centre
for Supercomputing (GCS) and the John von Neumann-Institut für
Computing (NIC) by providing computing time via the projects HMZ21, HMZ23
and HINTSPEC at JSC, as well as projects GCS-HQCD and GCS-MCF300 
at HLRS and LRZ. We also gratefully acknowledge the scientific 
support and HPC resources provided by NHR-SW of Johannes 
Gutenberg-Universität Mainz (project NHR-Gitter).
This work has been supported by Deutsche Forschungsgemeinschaft
(German Research Foundation, DFG) through the Collaborative Research 
Center 1660 “Hadrons and Nuclei as Discovery Tools”, under grant HI~2048/1-2
(Project No.\ 399400745), and through the Cluster of Excellence
``Precision Physics, Fundamental Interactions and Structure of
Matter'' (PRISMA+ EXC 2118/1), funded within the German Excellence
strategy (Project No.\ 390831469). This project has received funding
from the European Union's Horizon Europe research and innovation
programme under the Marie Sk\l{}odowska-Curie grant agreement
No.\ 101106243. 
A.R. was supported by the programme Netzwerke 2021, an initiative of the Ministry of Culture and Science of the State of Northrhine Westphalia, in the NRW-FAIR network, funding code NW21-024-A.
\normalsize

\bibliographystyle{JHEP}
\bibliography{biblio}

\end{document}